\documentclass[journal]{IEEEtran}

\usepackage{myStyleIEEE}
\usepackage{nomencl}
\IEEEoverridecommandlockouts

\usepackage{etoolbox}
\renewcommand\nomgroup[1]{%
  \item[\bfseries
  \ifstrequal{#1}{P}{A. Parameters}{%
  \ifstrequal{#1}{V}{C. Variables}{%
  \ifstrequal{#1}{S}{B. Sets and Indices}{}}}%
]}

\begin{document}
\bstctlcite{IEEEexample:BSTcontrol}

\title{Frequency-Constrained Resilient Scheduling of Microgrid: A Distributionally Robust Approach}

\newtheorem{proposition}{Proposition}
\renewcommand{\theenumi}{\alph{enumi}}

\newcommand{\Zhongda}[1]{\textcolor{magenta}{$\xrightarrow[]{\text{Zhongda}}$ #1}}

\author{Zhongda~Chu,~\IEEEmembership{Student~Member,~IEEE,}
 Ning~Zhang,~\IEEEmembership{Senior Member,~IEEE} 
        and Fei~Teng,~\IEEEmembership{Senior Member,~IEEE} 
   \thanks{

Zhongda Chu and Fei Teng (Corresponding author) are with Department of Electrical and Electronic Engineering, Imperial College London, U.K. Ning Zhang is with Department of Electrical Engineering, Tsinghua University, China.  
}
     
\vspace{-0.5cm}}
\maketitle
\IEEEpeerreviewmaketitle

\begin{abstract}
In order to prevent the potential frequency instability due to the high Power Electronics (PE) penetration under an unintentional islanding event, this paper presents a novel microgrid scheduling model which explicitly models the system frequency dynamics as well as the long/short term uncertainty associated with renewable energy resources and load. Synthetic Inertia (SI) control is applied to regulating the active power output of the Inverter-Based Generators (IBGs) to support the post-islanding frequency evaluation. The uncertainty associated with the noncritical load shedding is explicitly modeled based on the distributionally robust formulation to ensure resilient operation during islanding events. The resulted frequency constraints are derived analytically and reformulated into Second-Order Cone (SOC) form, which are further incorporated into the microgrid scheduling model, enabling optimal SI provision of Renewable Energy Sources (RESs) from the micorgrid perspective. With the SOC relaxation of the AC power flow constraints, the overall problem is constructed as a mixed-integer SOC Programming (MISOCP). The effectiveness of the proposed model is demonstrated based on modified IEEE 14-bus system.
\end{abstract}


\makenomenclature
\renewcommand{\nomname}{List of Symbols}
\mbox{}
\nomenclature[P]{$\bar P_{b}^{\mathrm{ch}}$/$\bar P_{b}^{\mathrm{dch}}$}{Maximum charging/discharging rate of storage unit b$\,[\mathrm{MW}]$}
\nomenclature[P]{$\Delta f_{\mathrm{lim}}$}{Limit of frequency deviation$\,[\mathrm{Hz}]$}
\nomenclature[P]{$f_0$}{Nominal frequency$\,[\mathrm{Hz}]$}
\nomenclature[P]{$\Delta \dot f_{\mathrm{lim}}$}{Limit of rate of change of frequency$\,[\mathrm{Hz/s}]$}
\nomenclature[P]{$E_c$}{Energy capacity of storage units$\,[\mathrm{MWh}]$}
\nomenclature[P]{$T_s$}{Time interval of constant power provision from storage units$\,[\mathrm{s}]$}
\nomenclature[P]{$D_0$}{Load-dependent damping$\,[\mathrm{MW/Hz}]$}
\nomenclature[P]{$H_g$}{Inertia time constant of SG g$\,[\mathrm{s}]$}
\nomenclature[P]{$T_d$}{PFR delivery time $\,[\mathrm{s}]$}
\nomenclature[P]{$\alpha$}{Linear coefficient of $\sigma$}
\nomenclature[P]{$\Delta P_L^{\mathrm{max}}$}{Maximum loss of generation $\,[\mathrm{MW}]$}
\nomenclature[P]{$\eta$}{Confidence level for the nadir constraint}
\nomenclature[P]{$d$/$k$/$K$/$N$}{Constants for piece-wise liearization of \eqref{x1^2}}
\nomenclature[P]{$a_n$/$b_n$}{Linear coefficients for piece-wise liearization \eqref{x1^2}}
\nomenclature[P]{$M$/$M'$}{Constants for big M methods}
\nomenclature[P]{$\pi_s$}{Probability of scenario s}
\nomenclature[P]{$c^{SU}_g$}{Start-up cost of SGs}
\nomenclature[P]{$c^{R1}_g$/$c^{R2}_g$}{Running cost of fixed/flexible generators}
\nomenclature[P]{$c^{VOLL}$}{Value of lost load}
\nomenclature[P]{$Y_{i,sh}$/$Y_{j,sh}$}{Shunt susceptance of branch ij}
\nomenclature[P]{$V_{\mathrm{min},i}/V_{\mathrm{max},i}$}{Minimum/maximum permissible voltage of bus $i$}
\nomenclature[P]{$S_{\mathrm{max}}$}{Maximum permissible apparent power of branch ij}
\nomenclature[P]{$P_W^C$/$P_M^C$}{Total capacity of wind/PV generation}

\nomenclature[V]{$P_{b}$}{Normal output power of storage unit b$\,[\mathrm{MW}]$}
\nomenclature[V]{$H_{s_b}$}{Synthetic inertia from storage unit b$\,[\mathrm{MWs/Hz}]$}
\nomenclature[V]{$H_{s_w}$}{Synthetic inertia from WT w$\,[\mathrm{MWs/Hz}]$}
\nomenclature[V]{$\Delta f$}{Microgrid frequency deviation$\,[\mathrm{Hz}]$}
\nomenclature[V]{$\Delta \dot f$}{Rate of change of frequency$\,[\mathrm{Hz/s}]$}
\nomenclature[V]{$\Delta P_{C}$}{Constant power provision after nadir from storage units$\,[\mathrm{MW}]$}
\nomenclature[V]{$\mathrm{SoC}$}{State of charge of storage units}
\nomenclature[V]{$\Delta P_w$}{Additional output power from WTs during frequency events$\,[\mathrm{MW}]$}
\nomenclature[V]{$\Delta \Tilde P_a$}{Output power of MPE$\,[\mathrm{MW}]$}
\nomenclature[V]{$D_s$}{Damping reduction due to the WT SI provision$\,[\mathrm{MW/Hz}]$}
\nomenclature[V]{$\gamma$}{Coefficient for damping reduction approximation$\,[\mathrm{MW/Hz}]$}
\nomenclature[V]{$H_c$}{Total inertia from conventional SG$\,[\mathrm{MWs/Hz}]$}
\nomenclature[V]{$H_r$}{Total synthetic inertia$\,[\mathrm{MWs/Hz}]$}
\nomenclature[V]{$H$}{Overall microgrid inertia$\,[\mathrm{MWs/Hz}]$}
\nomenclature[V]{$D$}{Overall microgrid damping$\,[\mathrm{MW/Hz}]$}
\nomenclature[V]{$\Delta R$}{Total PFR from SGs$\,[\mathrm{MW}]$}
\nomenclature[V]{$R$}{Total PFR delivered by time $T_d$$\,[\mathrm{MW}]$}
\nomenclature[V]{$\Delta P_{L_0}$}{Loss of generation due to the islanding event$\,[\mathrm{MW}]$}
\nomenclature[V]{$\Delta P_{L}$}{Equivalent loss of generation$\,[\mathrm{MW}]$}
\nomenclature[V]{$\Delta P_{D}$}{Noncritical load shedding$\,[\mathrm{MW}]$}
\nomenclature[V]{$\Delta P_{L_\mu}$}{Mean of $\Delta P_L$ $\,[\mathrm{MW}]$}
\nomenclature[V]{$\Delta P_{D_\mu}$}{Mean of $\Delta P_D$ $\,[\mathrm{MW}]$}
\nomenclature[V]{$\sigma$}{Standard deviation of $\Delta P_D$ and $\Delta P_L$ $\,[\mathrm{MW}]$}
\nomenclature[V]{$x_1$/$x_2$}{Ancillary variables for nadir constraint reformulation}
\nomenclature[V]{$z_n$/$z_{n_{1,2}}$}{Binary variables for $x_2$ interval indication}
\nomenclature[V]{$z_{t,s,g}$}{Start up binary variable of generator $g$ at time step $t$ in scenario $s$}
\nomenclature[V]{$y_{t,s,g}$}{Operation status of generator $g$ at time step $t$ in scenario $s$}
\nomenclature[V]{$p_{t,s,g}$}{Active power produced by generator $g$ at time step $t$ in scenario $s$ $\,[\mathrm{MW}]$}
\nomenclature[V]{$p_{t,s,l}^c$/$q_{t,s,l}^c$}{Active/reactive load shedding at load $l$ at time step $t$ in scenario $s$}
\nomenclature[V]{$p_{t,s,ij}^c$/$q_{t,s,ij}^c$}{Active/reactive power flow from bus $i$ to $j$}
\nomenclature[V]{$P_{t,s,im}^c$/$Q_{t,s,im}^c$}{Imported power from the main grid}
\nomenclature[V]{$V_{t,s,i}$}{Voltage of bus $i$ at time step $t$ in scenario $s$}

\nomenclature[S]{$b\in\mathcal{B}$}{Set of energy storage units}
\nomenclature[S]{$w\in\mathcal{W}$}{Set of Wind generation units}
\nomenclature[S]{$m\in\mathcal{M}$}{Set of PV generation units}
\nomenclature[S]{$t\in\mathcal{T}$}{Set of time steps in the optimization}
\nomenclature[S]{$g\in\mathcal{G}$}{Set of conventional SGs}
\nomenclature[S]{$l\in\mathcal{L}$}{Set of loads}
\nomenclature[S]{$g\in\mathcal{G}_1$}{Set of fast SGs}
\nomenclature[S]{$g\in\mathcal{G}_2$}{Set of slow SGs}
\nomenclature[S]{$\mathbf{D}\in\mathcal{P}$}{Ambiguity set of $\Delta P_L$}
\nomenclature[S]{$n\in\mathcal{N} \cup\{N\} $}{Set of piece-wise linear segments}
\nomenclature[S]{$s\in\mathcal{S}$}{Set of scenarios}
\nomenclature[S]{$\theta \in \Omega_{\theta -i}$}{Set of units connected to bus i, $\theta\in \{ g, w, m, b, l\}$}
\nomenclature[S]{$ij\in\mathcal{R} $}{Set of branches}
\printnomenclature

\section{Introduction} \label{sec:1}
Microgrids, distribution systems integrated with large scale of RESs, storage devices and controllable loads have been a promising concept for reliable and flexible electricity supply in an environment-friendly manner \cite{Agrawal2011MICROGT}. They are connected to the main grid at the Point of Common Coupling (PCC), providing the capability of power transmission in both directions. Microgrids can operate in islanded mode by disconnecting itself from the main grid when subjected to external disturbances, making them highly beneficial to customers and utilities \cite{7420802}.

Due to the resiliency benefits of microgrids, extensive research has been conducted on microgrid scheduling aiming to achieve optimal microgrid operation and management \cite{NOSRATABADI2017341,7120901}. A stochastic microgrid scheduling model is proposed in \cite{6662462} to address the intermittency and variability of the RESs. Applying the chance constrained approach, the authors in \cite{7438879} formulate the grid-connected microgrid scheduling problem as linear programming. The studies in \cite{8254387} develop a distributionally robust chance-constrained energy management for islanded microgrids with the uncertainty of renewable generation captured through a novel ambiguity set. \cite{6839110} presents a resiliency-oriented microgrid optimal scheduling model in islanded operating mode to minimize the microgrid load curtailment. The demand response from Electric Vehicles (EV) is incorporated into the islanded microgrid scheduling problem to minimize the operating and EV charging costs \cite{8807105}. 

Most of the literature in this vein focuses on the microgrid scheduling in either grid-connected or islanded mode with less attention being paid on the influence of the transition between the two modes. The islanding events of microgrids force its exchanged power with the main grid to zero, resulting in power unbalance between generation and demand. Furthermore, due to the high PE penetration in microgrids, this unbalanced power can lead to large frequency deviations, even blackout and system collapse. 

The authors in \cite{6671442} consider multi-period islanding constraints in a centralized microgrid optimal scheduling model. The solution is examined for islanding to ensure the microgrid has sufficient online capacity for quickly switching to the islanded mode on request. \cite{LIU2017197,6615946} propose an optimal scheduling strategy for microgrid operation considering constraints of islanding capability. Probability of Successful Islanding (PSI) is introduced as a new concept to ensure there is enough reserve to cover the load after islanding events. Similarly, in \cite{8974607}, a new optimal strategy for scheduling of reconfigurable microgrids is presented while maintaining the PSI above a certain level. Considering the uncertainty of RESs and demand, the microgrid scheduling is formulated as a chance-constrained global optimization problem. 

However, the above methods only focus on determining the feasible region of the spinning reserve to guarantee the PSI while neglecting the detailed frequency dynamics and the frequency support from IBGs. It is improved by \cite{8561183} where a comprehensive optimization and real-time control framework for maintaining frequency stability of multi-microgrid networks under islanding events are proposed. The frequency dynamics and SI from IBGs are also considered leading to highly nonlinear frequency constraints. An iterative algorithm is developed based on the cutting plan approach to incorporate the post-contingency frequency constraints, which increases its implementing complexity. Deep learning is applied in \cite{9165193} to approximate the nonlinear nadir constraint using a neural network such that an MILP-based microgrid scheduling problem can be formulated. Nevertheless, the detailed SI modeling from IBGs is not considered and the uncertainty associated with the noncritical load shedding due to the forecasting errors and the relays has not been discussed. 

Motivated by previous research, this paper proposes a novel optimal microgrid scheduling model considering the state-of-art SI control scheme from PV-storage system and Wind Turbines (WTs) to minimize the microgrid operation cost while ensuring the frequency security after islanding events. The contributions of this paper are summarized as follow:
\begin{itemize}
    \item A novel microgrid scheduling model is proposed, which optimizes microgrid operating conditions, noncritical load shedding as well as the SI from IBGs such that the frequency constraints after microgrid islanding events can be maintained.
    \item A distributionally robust approach is adopted to account for the uncertainty associated with noncritical load shedding leading to distributionally robust chance constraint formulation of the frequency metrics.
    \item The highly nonlinear frequency constraints are further effectively reformulated into SOC form and incorporated into the microgrid scheduling model, resulting in an overall MISOCP together with the SOC approximation of the AC power flow.
\end{itemize}
The rest of this paper is structured as follows. Section \ref{sec:2} introduces the SI control modeling of RESs and the overall microgrid frequency dynamics, based on which the analytical expressions of the frequency metrics are derived. The uncertainty associated with the noncritical load shedding is discussed in Section \ref{sec:3}, leading to distributionally robust frequency constraints and the nonlinear nadir constraint is further reformulated into SOC form. The overall MISOCP-based microgrid scheduling model is presented in Section \ref{sec:4}, followed by case studies (Section \ref{sec:5}) illustrating the value and performance of the proposed model. Finally, section \ref{sec:6} concludes the paper.

\section{Frequency Dynamics of Microgrid with SI Provision from RESs} \label{sec:2}
In this section, the microgrid frequency dynamics are investigated. The Rate of Change of Frequency (RoCoF), frequency nadir and steady-state value subsequent to islanding events are derived considering the provision of SI from RESs. {Due to the low efficiency during normal operation, the deloading control strategy is not considered in this paper. Instead,} in order to maximize the energy captured by PV systems and WTs, we assume that they are controlled through MPPT strategy during normal operation. For PV systems, energy storage devices are used to provide frequency support during the system disturbance, whereas stored kinetic energy is utilized in the WTs. {It is the synthetic inertia rather than the conventional droop control that is applied in the proposed model to provide higher power injection at the beginning of islanding events.} It should be noted that only loss of generation (or increase of load) is considered in this paper since the opposite situation can be easily dealt with by shifting the operating point of the RESs away from the optimal power point.

\subsection{Modeling of Frequency Support from Energy Storage Devices} \label{sec:2.1}
A state-of-the-art VSC control scheme previously described in \cite{8579100} is adapted to provide constant synthetic inertia during a disturbance. Furthermore, since the power injection associated with synthetic inertia approaches zero at the frequency nadir and has no impact on the steady-state frequency, constant power can be injected into the grid after the frequency nadir to improve the steady-state frequency if needed. The active power setpoint is adjusted through the outer control loop to achieve the desired power injection to the grid. Note that although it is possible to achieve more sophisticated control laws, e.g., adaptive virtual synchronous machine or online model predictive control, it would make the scheduling model much more complicated, thus not being considered in this paper. 

In order to determine an optimal and feasible frequency support provision during the scheduling process, it is essential to consider the limitation of both instantaneous power and total energy of the energy storage devices.

The instantaneous output power of the energy storage device $b\in\mathcal{B}$ during an entire frequency event, $\mathcal{T}_0$ is confined by the maximum charging/discharging rate, $\bar P_{b}^{\mathrm{ch}}$/$\bar P_{b}^{\mathrm{dch}}$:
\begin{equation}
    \label{P_HD}
    \bar P_{b}^{\mathrm{ch}} \le P_b - 2H_{s_b} \Delta \dot f(t) \le \bar P_{b}^{\mathrm{dch}},\,\,\;\forall t\in \mathcal{T}_0
\end{equation}
where $P_b$ and $H_{s_b}$ are the normal output power and synthetic inertia from the storage unit $b$. Equation \eqref{P_HD} is equivalent to:
\begin{equation}
    \label{P_HD_max}
    \bar P_{b}^{\mathrm{ch}} \le P_b + \max_{t\in \mathcal{T}_0}  \Big\{2H_{s_b} \left| \Delta \dot f(t)\right| \Big\} \le \bar P_{b}^{\mathrm{dch}}.
\end{equation}
Since $\left|\Delta \dot f(t)\right|$ is constrained by the RoCoF limit:
\begin{equation}
    0 \le \left|\Delta \dot f(t)\right| \le \Delta \dot f(0) \le \Delta \dot f_{\mathrm{lim}} ,
\end{equation}
\eqref{P_HD_max} can be rewritten in linear form as follows:
\begin{equation}
\label{Hv_lim}
    \bar P_{b}^{\mathrm{ch}} \le P_b + 2H_{s_b} \Delta \dot f_{\mathrm{lim}} \le \bar P_{b}^{\mathrm{dch}}.
\end{equation}
The energy required by synthetic inertia provision is negligible due to the small time scale of the inertial response. Hence, it is not considered as a constraint here.

Similarly, the limitation for the constant power provision $\Delta P_C$ after the frequency nadir can be formulated as:
\begin{subequations}
\label{energy_constr}
    \begin{align}
    \label{Pc_lim}
        \bar P_{b}^{\mathrm{ch}} \le P_b + \Delta P_C & \le \bar P_{b}^{\mathrm{dch}} \\
        \Delta P_C T_s & \le  \mathrm{SoC}_b \cdot E_{c,b}
    \end{align}    
\end{subequations}
where $T_s$ is the time interval of the constant power provision; $E_{c}$ and $\mathrm{SoC}$ are the energy capacity and the state of charge of the storage device. {Note that the above frequency support model and the operational constraints can also be applied to other microgrid battery storage units that are not connected with PV systems. In addition to the synthetic inertia provision, the energy storage devices are responsible to smooth the renewable generation fluctuation in shorter timescale and balance the inconsistent profiles between the load and generation in longer timescale (storage the energy when the generation is more than the demand and release the energy otherwise). All these services are simultaneously optimized during the microgrid scheduling process in our proposed model, such that the storage capacity can be optimally allocated in real time.}

\subsection{Modeling of Frequency Support from WTs} \label{sec:2.2}
The control framework proposed in \cite{9066910} is applied to provide optimal synthetic inertia from WTs. In the proposed model, active power is extracted from the stored kinetic energy of WTs to facilitate the frequency evaluation during the disturbance. Due to the complexity caused by incorporating the WT mechanical dynamics into the control design and the restriction of the stored kinetic energy, only short-term inertial response is provided from WTs.

\begin{figure}[!b]
    \centering
    \vspace{-0.35cm}
	\scalebox{0.735}{\includegraphics[]{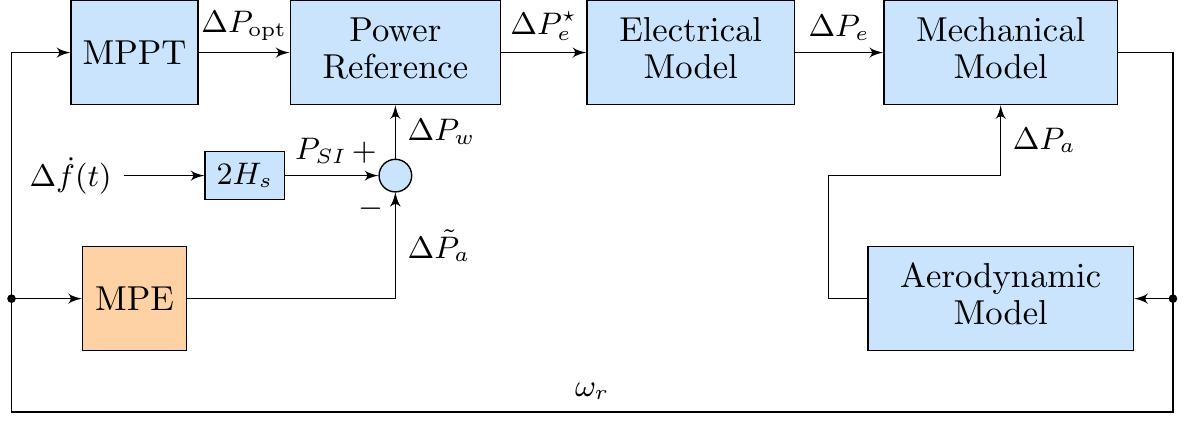}}
    \caption{\label{fig:Block}Block diagram of WT synthetic inertia control scheme.}
    \vspace{-0.35cm}
\end{figure}

Furthermore, the secondary frequency dip associated with the rotor speed deviation from the optimal operation point before the disturbance can be eliminated. This is achieved by adding a Mechanical Power Estimator (MPE) in the active power control loop of the WT grid-side converter as shown in Fig.~\ref{fig:Block}. As a result, the additional output power from WTs ($\Delta P_w$) during a system disturbance is the sum of synthetic inertia power $P_{SI}$ and the output of the MPE $\Delta \Tilde P_{a}$:
\begin{equation}
    \label{P_w}
    \Delta P_w = - 2 H_{s} \Delta \dot f + \Delta \Tilde P_{a}.
\end{equation}
In order to incorporate \eqref{P_w} into the system frequency dynamics, the highly nonlinear expression of $\Delta \Tilde P_{a}$ is further approximated by a negative system damping term:
\begin{equation}
    \label{WF_D}
    \Delta \Tilde P_{a} = D_s \Delta f = \gamma H_{s}^2 \Delta f.
\end{equation}
In addition, the total available synthetic inertia form WTs in the system can be estimated given the wind speed distribution as proposed in \cite{9066910}, \textcolor{black}{where the feasibility of frequency support from WTs and detailed control performance can be found as well}.

\subsection{Frequency Evaluation under Islanding Events}\label{sec:2.3}
Without the frequency support from the RESs, the frequency dynamics in a multi-machine microgrid can be expressed in the form of a single swing equation, under the premise of the Centre-of-Inertia (CoI) model \cite{7833096}:
\begin{equation}
    \label{sw1}
    2H_c\frac{\partial\Delta f(t)}{\partial t} = -D_0 \Delta f(t) + \Delta R(t) -\underbrace{(\Delta P_{L_0}-\Delta P_D)}_{\Delta P_L},
\end{equation}
where $\Delta P_{L_0}$, the loss of generation due to the islanding event at $t=0$ is a decision variable and can be viewed as a step disturbance. $\Delta P_D$ is the noncritical load shedding in order to maintain the post-contingency frequency within the limits, {which is a common practice in microgrid after islanding events.} It can be deferred or curtailed in response to economic incentives or islanding requirements. \textcolor{black}{Furthermore, $\Delta P_D$ is modeled as a decision variable with uncertainty and its mean ($\Delta P_{D_\mu}$) and standard deviation ($\sigma$) are assumed to be partially known.} More details regarding to the uncertainty of $\Delta P_D$ are discussed in Section \ref{sec:3}. {$\Delta P_L$ is the equivalent loss of generation which is always positive, as there is no point to shed more load than the lost power.}  Moreover, the PFR $\Delta R(t)$ from conventional Synchronous Generators (SGs) can be represented according to the following scheme \cite{6714513}:
\begin{equation}
\label{R}
\Delta R(t)=
     \begin{cases}
       \frac{R}{T_d}t &, \; 0\le t< T_d \\ 
       R &, \; T_d\le t
     \end{cases}
\end{equation}
with $T_d$ being the PFR delivered time and R being the total PFR delivered by time $T_d$; The total inertia of SGs is computed as:
\begin{equation}
    \label{H1}
    H_c =\frac{\sum_{g\in \mathcal{G}} H_g  P_g^\mathrm{max} y_g}{f_0}.
\end{equation}
Incorporate the frequency support from RESs as described in Section \ref{sec:2.1} and \ref{sec:2.2} into \eqref{sw1} leading to:
\begin{align}
    \label{sw2}
    & 2 \textcolor{black}{ \underbrace{\left(H_c + \sum_{b\in \mathcal{B}} H_{s_b} + \sum_{w\in \mathcal{W}} H_{s_w}\right)}_H} \frac{\partial\Delta f(t)}{\partial t}\\ &
    \nonumber
    = - \textcolor{black}{ \underbrace{\left(D_0 - \sum_{w\in \mathcal{W}} \gamma_w H_{s_w}^2\right)}_D} \Delta f(t) + \Delta R(t) -\Delta P_L
\end{align}
\textcolor{black}{where $H$ and $D$ are the overall system inertia and damping respectively; $b\in \mathcal{B}$ and $w\in \mathcal{W}$ are the set of energy storage units and wind generation units. Note that the inertia in the system is now the combination of SGs' ($H_c$) and the SI from the energy storage devices ($H_{s_b}$) and WTs ($H_{s_w}$). The system damping is decreased by $\sum_{w\in \mathcal{W}} \gamma_w H_{s_w}^2$ due to the SI provision from WTs. It represents the the side effect of SI provision from wind turbines through overproduction scheme, i.e.,  the output power reduction due to the deviation from the optimal operating point \cite{9066910}.}

Based on the frequency dynamics, the analytical expression of the maximum instantaneous RoCoF $(\Delta \dot f_\mathrm{max}\equiv\Delta \dot f|_{t=0^+})$ is identified as:
\begin{equation}
\label{rocof}
    \Delta \dot f|_{t=0^+} = -\frac{\Delta P_L}{2H}.
\end{equation}
It can be maintained within the RoCoF limits by choosing an appropriate system inertia $H$. Solving the differential equation \eqref{sw2} gives the microgrid frequency evaluation during an islanding event:
\begin{equation}
\label{f(t)}
    \Delta f(t) = \left(\frac{\Delta P_L}{D}+\frac{2HR}{T_d D^2}\right)\left(e^{-\frac{D}{2H}t}-1\right) + \frac{R}{T_d D}t,
\end{equation}
valid $\forall t\in [0,t_n]$. The time instant $t_n$ of frequency nadir is derived by setting the derivative of \eqref{f(t)} to zero:
\begin{equation}
\label{tn}
    \Delta \dot f(t_n)=0 \longmapsto t_n = \frac{2H}{D}\ln{\left(\frac{T_d D \Delta P_L}{2HR}+1\right)}.
\end{equation}
Substituting \eqref{tn} into \eqref{f(t)} leads to the expression for frequency nadir $(\Delta f_\mathrm{max}\equiv\Delta f(t_n))$:
\begin{equation}
\label{nadir}
    \Delta f(t_n) = \frac{2HR}{T_d D^2} \ln{\left(\frac{T_d D \Delta P_L}{2HR}+1\right)}-\frac{\Delta P_L}{D}.
\end{equation}
It is understandable that the dependence of the frequency nadir on $H$, $D$, and $\Delta P_L$ through a highly-nonlinear relationship makes it difficult to be incorporated into the microgrid scheduling model. A SOC reformulation is proposed to cope with this problem as demonstrated in Section \ref{sec:3.1}.

It should be noted that in order to derive the analytical expressions of maximum RoCoF and frequency nadir, only the inertial response from RESs are incorporated in \eqref{sw2}. However, when deriving the steady-state frequency $(\Delta f_\mathrm{max}^\mathrm{ss}\equiv\Delta f|_{t=\infty})$, the constant power injection $\Delta P_C$ from energy storage devices needs to be considered:
\begin{equation}
\label{fss}
    \Delta f|_{t=\infty} = \frac{R+\Delta P_C-\Delta P_L}{D}.
\end{equation}
{Note that the secondary frequency response from conventional generators and the associated frequency restoration process after steady-state are not considered in this paper.} 

Having obtained the analytical expressions of the frequency metrics during an islanding event \eqref{rocof} \eqref{nadir} and \eqref{fss}, they should be bounded within predescribed limits in the microgrid scheduling model by selecting proper $H$, $R$, $\Delta P_C$ and $\Delta P_L$. These quantities can all be decided deterministically except the equivalent loss of generation $\Delta P_L$ due to the uncertainty associated with the noncritical load shedding $\Delta P_D$. At the beginning of an islanding event, the noncritical load is disconnected from the microgrid in order to support the frequency evaluation. However, the exact value of the shed load is unknown during the scheduling period, thus increasing the complexity of the scheduling problem as elaborated in the next section.

\section{Distributionally Robust Chance Constraints of Frequency Metrics}\label{sec:3}
{The uncertainty associated with the noncritical load shedding $\Delta P_D$ stems from different aspects. On the one hand, forecasting error always exists during the microgrid scheduling process in terms of the actual demand and the noncritial load percentage. On the other hand,} depending on the specific load shedding strategies\cite{970031,BAKAR2017161}, the uncertainty level of $\Delta P_D$ varies. Traditionally, the practice setting of the RoCoF and frequency relays are mainly based on the experts' experiment \cite{5697619}. More advanced noncritical load control schemes have also been proposed to mitigate frequency variations during islanding events. {For instance, the emergence demand response has been considered in the setting of the under frequency load shedding relays and the design of the load shedding schemes \cite{RAFINIA2020102168,6213578,7822924,8012432}.} Therefore, it is complicated to derive the detailed distribution of $\Delta P_D$ through either model-based or data-driven approaches at microgrid scheduling stage. {As a result, the equivalent loss of generation $\Delta P_L$ as defined in \eqref{sw1} also presents uncertainty of the same level.} In order to account for the uncertainty of $\Delta P_L$, the frequency constraints are reformulated through distributionally robust optimization. Assume the first- and second-order moments of $\Delta P_L$ are {decision-dependent} whereas the exact probability distribution $\mathbf{D}$ is unknown. This is modeled by the following ambiguity set:
\begin{align}
\label{N(P_L)}
    \mathcal{P} =\Big\{\mathbf{D} \in \Phi(\Delta P_L):\; & \mathbb{E}^\mathbf{D}(\Delta P_L) = \Delta P_{L_{\mu}},\nonumber \\
    &\mathrm{Var}^\mathbf{D}(\Delta P_L) =\sigma^2 \Big\}
\end{align}
where $\Phi(\cdot)$ is the probability density function; $\Delta P_{L_{\mu}}=\Delta P_{L_0}-\Delta P_{D_\mu}$ and $\sigma$ denote the mean and standard deviation of the equivalent loss of generation given the distribution $\mathbf{D}$. {Since the more noncritical load is about to be shed, the higher uncertainty level it presents,} it is reasonable to assumed that $\sigma$ depends on the decision variable $\Delta P_{D_\mu}$ with a linear coefficient $\alpha$, i.e.,
\begin{equation}
\label{alpha}
    \sigma= \alpha \Delta P_{D_\mu}.
\end{equation}
{Having defined the mean and standard deviation of $\Delta P_D$ and $\Delta P_L$, it should be clear now that the only decision to be made associated with noncritical load shedding is $\Delta P_{D_{\mu}}$, meaning that statistically the mean of noncritical load shedding equals to the decision made by the system operator. However, for the realization in a single islanding event, it is very likely for $\Delta P_D$ to deviate from the decision variable $\Delta P_{D_\mu}$ characterized by its standard deviation $\sigma$.}

\subsection{Nadir Constraint Reformulation} \label{sec:3.1}
Based on the method proposed in \cite{9066910}, the nadir constraint $\Delta f(t_n)\le \Delta f_{\mathrm{lim}}$ can be converted into the following nonlinear from:
\begin{equation}
\label{nadir_c_D0}
    HR\ge \frac{\Delta P_L^2T_d}{4\Delta f_\mathrm{lim}}-\frac{\Delta P_L T_d D_0}{4}+ \frac{\Delta P_L T_d \sum_{w\in \mathcal{W}} \gamma_w H_{s_w}^2}{4}.
\end{equation}
Since $\sum_{w\in \mathcal{W}} \gamma_w H_{s_w}^2$ is much less than $D_0$, the $\Delta P_L$ in the last term of \eqref{nadir_c_D0} is set to be a constant ($\Delta P_L^{\mathrm{max}}$ for conservativeness). As a result, \eqref{nadir_c_D0} can be rewritten as follows:
\begin{equation}
\label{nadir_c1_D0}
    HR - \underbrace{\frac{\Delta P_L^{\mathrm{max}} T_d  \sum_{w\in \mathcal{W}} \gamma_w H_{s_w}^2}{4}}_c\ge \frac{T_d}{4\Delta f_\mathrm{lim}} \Delta P_L^2 -\frac{T_d D_0}{4}\Delta P_L.
\end{equation}
Equation \eqref{nadir_c1_D0} can be viewed as a quadratic inequality of $\Delta P_L$. Since $T_d/(4\Delta f_\mathrm{lim})>0$, it is equivalent to:
\begin{align}
\label{PL_nadir}
    \Delta P_L \in \Bigg[ & \underbrace{\frac{D_0 \Delta f_\mathrm{lim}}{2} -\frac{\Delta f_\mathrm{lim}}{T_d}\sqrt{\frac{T_d^2 D_0^2}{4}+\frac{4 T_d}{\Delta f_\mathrm{lim}} (HR-c)}}_{\Delta \underline {P}_L}, \nonumber \\ 
    & \underbrace{\frac{D_0 \Delta f_\mathrm{lim}}{2}  + \frac{ \Delta f_\mathrm{lim}}{T_d}\sqrt{\frac{T_d^2 D_0^2}{4}+\frac{4 T_d}{\Delta f_\mathrm{lim}} (HR-c)}}_{\Delta \bar{P}_L} \Bigg]
\end{align}

Therefore, the distributionally robust nadir chance constraint is formulated as follows:
\begin{equation}
\label{Pr_nadir}
    \min_{\mathbf{D}\in\mathcal{P}} \mathrm{Pr} \Big\{ \Delta P_L \in \big[\Delta  \underline{P}_L, \Delta \bar{P}_L \big]   \Big\} \ge \eta.
\end{equation}
Here only the cases of generation loss are considered, i.e., $\Delta P_L > 0$, therefore, the probability of $\Delta P_L$ being negative is zero. Moreover, it can be derived from \eqref{PL_nadir} that $\Delta \underline {P}_L < 0$ always holds. Hence, \eqref{Pr_nadir} is equivalent to:
\begin{equation}
\label{Pr_nadir_1}
    \min_{\mathbf{D}\in\mathcal{P}} \mathrm{Pr} \Big\{ \Delta P_L \le \Delta \bar{P}_L \Big\} \ge \eta.
\end{equation}
By applying Chebyshev inequality \cite{Calafiore2006}, the nonconvex constraint \eqref{Pr_nadir_1} can be reformulated as follows:
\begin{equation}
\label{nadir_nonlinear}
    \Delta \bar{P}_L \ge \Delta P_{L_{\mu}}+ \underbrace{\sqrt{\frac{\eta}{1-\eta}}}_{\xi} \sigma.
\end{equation}
Substituting the expression of $\Delta \bar{P}_L$ into \eqref{nadir_nonlinear} yields:
\begin{align}
    \label{nadir_nonlinear_2}
    HR\ge & \frac{T_d}{4} \left[   \frac{(\Delta P_{L_{\mu}}+\xi \sigma)^2}{\Delta f_\mathrm{lim}}  - D_0( \Delta P_{L_{\mu}}+\xi \sigma) \right] \nonumber    \\
    + & \frac{\Delta P_L^{\mathrm{max}} T_d  \sum_{w\in \mathcal{W}} \gamma_w H_{s_w}^2}{4}.
\end{align}
Introduce ancillary variables $x_1$ such that:
\begin{align}
    x_1^2 & =\frac{(\Delta P_{L_{\mu}}+\xi \sigma)^2}{\Delta f_\mathrm{lim}}  - D_0( \Delta P_{L_{\mu}}+\xi \sigma) \nonumber \\
    \label{x1^2}
    & = \underbrace{\frac{\Delta P_{L_{\mu}}+\xi \sigma}{\sqrt{\Delta f_\mathrm{lim}}}}_{x_2} \left( \frac{\Delta P_{L_{\mu}}+\xi \sigma}{\sqrt{\Delta f_\mathrm{lim}}}- \underbrace{\sqrt{\Delta f_\mathrm{lim}} D_0}_d \right).
\end{align}
It can be proved that $x_2\ge d$ always holds given a small system damping. Hence, \eqref{x1^2} is a well-defined real value constraint. The nadir constraint \eqref{nadir_nonlinear_2} can be thus rewritten as a SOC form:
\begin{equation}
\label{nadir_soc}
    HR \ge \frac{T_d}{4} x_1^2 + \frac{\Delta P_L^{\mathrm{max}} T_d  \sum_{w\in \mathcal{W}} \gamma_w H_{s_w}^2}{4}
\end{equation}
However, the nonconvex constraint \eqref{x1^2} can not be included in the MISOC optimization directly. A set of linear constraints are used to approximate this relationship conservatively. Depending on the ratio of $x_2$ to $d$, the relationship between $x_1$ and $x_2$ described by \eqref{x1^2} can be piece-wise characterized by $N$ linear expression:
\begin{subequations}
\label{nadir_linear}
    \begin{align}
    x_1 = a_n & x_2 +b_n D_0, \,\,\forall n \in \mathcal{N} \nonumber\\
    & if\,\; k (n-1) +1\le \frac{x_2}{d} < k n+1\\
    \label{asymp}
    x_1 = a_N & x_2 +b_N D_0, \nonumber\\
    & if\,\; K+1  \le \frac{x_2}{d}
\end{align}
\end{subequations}
where $k$ defines the step size of $x_2$ in the first $N-1$ linear constraints, i.e., $n\in \mathcal{N}=\{1,2,...,N-1\}$:
\begin{equation}
\label{step_k}
    k = \frac{K }{N-1}
\end{equation}
with $K$ being a predefined constant. Theoretically, the ratio of $x_2$ to $d$ can be vary large. To reduce the value of $N$, a single linearized constraint is applied if $x_2$ is large enough compared to $d$, i.e., $ K \le x_2/d$ as shown in \eqref{asymp}. The coefficient $a_n$ and $b_n$ can be calculated as:
\begin{subequations}
\begin{align}
    \label{an_bn}
    a_n & = \dv{x_1}{x_2}\Bigr|_{x_2=(kn+1)D_0}= \frac{2nk+1}{2\sqrt{n^2k^2+nk}},\,\,\forall n\in \mathcal{N}\\
    b_n &=\frac{-nk-1}{2\sqrt{n^2k^2+nk }}, \,\,\forall n\in \mathcal{N}\\
    a_N &= 1 \\
    b_{N} &= -0.5
\end{align}
\end{subequations}
However, the $N$ linear constraints defined in \eqref{nadir_linear} cannot be included in the optimization simultaneously. Instead, only one of them needs to hold while others should be relaxed depending on the relationship between $x_2$ and $d$. Therefore, $N$ binary variables, $z_n,\,\forall n\in \mathcal{N}\cup\{N\}$ are introduced to indicate to which interval $x_2$ belongs:
\begin{subequations}
    \begin{align}
    \label{z_n}
    z_{n\in\mathcal{N}} & = 
    \begin{cases}
        1 &if\,\,d k(n-1) \le x_2 < d k n \\
        0 & \mathrm{otherwise}.
    \end{cases}\\
    z_{N} & = 
    \begin{cases}
        1 &if\,\,K d \le x_2 \\
        0 & \mathrm{otherwise}.
    \end{cases}
    \end{align}
\end{subequations}
Equation \eqref{z_n} can be rewrite in the following form by defining ancillary binary variables $z_{n_1},\,z_{n_2},\,\forall n \in \mathcal{N}$:
\begin{subequations}
\label{z_n12}
    \begin{align}
    z_{n_1} & = 
    \begin{cases}
        1 &if\,\,d k(n-1) \le x_2 \\
        0 & \mathrm{otherwise}
    \end{cases}\\
    z_{n_2} & = 
    \begin{cases}
        1 &if\,\,x_2 < dkn \\
        0 & \mathrm{otherwise}
    \end{cases}\\
    z_n & = z_{n_1}  + z_{n_2} -1.
    \end{align}
\end{subequations}
As a result, the conditional constraints \eqref{z_n12} can be transformed into linear constraints $\forall n\in \mathcal{N}$:
\begin{subequations}
    \label{z_n12_linear}
    \begin{align}
    & 0 < x_2 - dkn + M z_{n_1} \le M
    \\
    & 0 \le d k(n-1) - x_2 + M z_{n_2} < M
    \\
    & z_n = z_{n_1}  + z_{n_2} -1 \\
    & z_{n_1},\, z_{n_2}\,\in \{0,1\}
    \end{align}
\end{subequations}
where $M>0$ is a sufficiently large constant. Similarly, linear constraints for $Z_N$ can be expressed as follows:
\begin{subequations}
    \label{z_N}
    \begin{align}
    & 0 < x_2 - K d+ M z_{N} \le M \\
    & z_{N}\,\in \{0,1\}
    \end{align}
\end{subequations}
Based on the interval indicators $z_n$, the equality constraints \eqref{nadir_linear} are relaxed as follows to ensure feasibility:
\begin{equation}
    \label{nadir_linear_1}
    x_1 \ge a_n x_2 +b_n d + (z_n-1) M', \;\;\;\; \forall n\in \mathcal{N}\cup\{N\}
\end{equation}
where $M'>0$ is a large enough constant. It should be noted that the equality in \eqref{nadir_linear_1} will be automatically obtained during the optimization process since $HR$ term in the nadir constraint \eqref{nadir_soc} is positively correlated to the objective function. As a result, the original distributionally robust nadir chance constraint \eqref{nadir_nonlinear_2} is now reformulated into the SOC form: \eqref{nadir_soc}\eqref{z_n12_linear}\eqref{z_N}\eqref{nadir_linear_1}.
\subsection{RoCoF and steady-state Constraints Reformulation}
Based on the expressions derived in \eqref{rocof} and \eqref{fss}, the distributionally robust frequency constraints of RoCoF and steady-state can be formulated as follows:
\begin{equation}
\label{rocof_cstr}
    2H \Delta \dot f_{\mathrm{lim}} \ge \Delta P_{L_\mu}+\xi\sigma
\end{equation}

\begin{equation}
\label{fss_cstr}
    R+\Delta P_C+ (D_0  - \sum_{w\in \mathcal{W}}\gamma_w H_{s_w}^2) \Delta  f^{ss}_{\mathrm{lim}} \ge \Delta P_{L_\mu}+\xi\sigma
\end{equation}
with $\dot f_{\mathrm{lim}}$ and $ \Delta  f^{ss}_{\mathrm{lim}} $ being the pre-specified limits for maximum instantaneous RoCoF and steady-state frequency deviation. The nonlinear term associated with $H_{s_w}^2$ can be effectively linearized as demonstrated in \cite{9066910}.
\section{MISOCP-based Microgrid Scheduling} \label{sec:4}
\begin{figure}[!t]
    \centering
    \vspace{-0.35cm}
	\scalebox{0.42}{\includegraphics[]{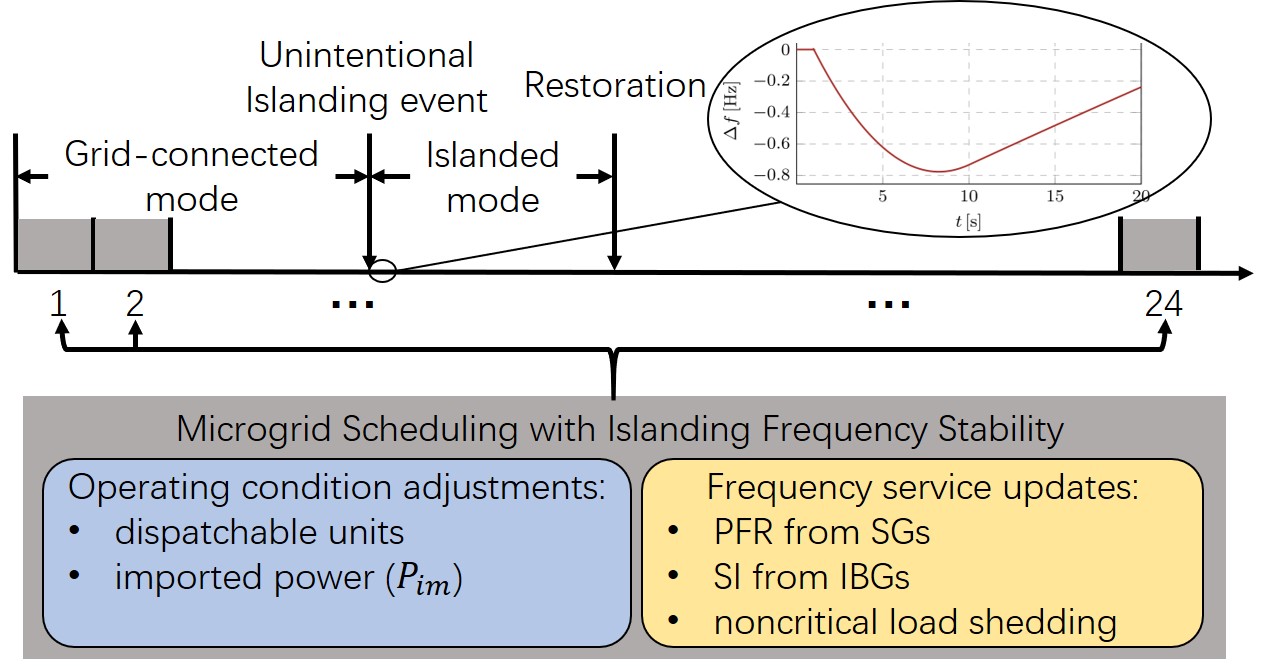}}
    \caption{\label{fig:timescale}{\textcolor{black}{Relationship between microgrid scheduling and frequency control.}}}
    \vspace{-0.35cm}
\end{figure}

In this section, a two-stage stochastic microgrid scheduling model is introduced to determine the optimal generator dispatch, wind/PV curtailment and load shedding with frequency security constraints. \textcolor{black}{The relationship between the microgrid scheduling problem and the frequency control is demonstrated through Fig.~\ref{fig:timescale}. During normal operation (grid-connected mode), the microgrid operates according to the results obtained from the scheduling problem with optimal operating conditions and the optimal frequency responses updated in each hour, as indicated by the blue and yellow areas respectively in the figure below. Notably, the commands related to the frequency services would only lead to the parameter updates in the associated controllers. Those services would not be triggered unless an islanding event is detected, which can be achieved by monitoring the main breaker at the PCC or measuring the RoCoF. Due to the frequency constraints in the microgrid scheduling model, the frequency security after an islanding event at any time can be guaranteed through a most cost-efficient way.}

Consider a microgrid with a set of SGs units $g\in \mathcal{G}$ and loads $l\in \mathcal{L}$. The generation units are further categorized into two groups, i.e. $\mathcal{G}=\mathcal{G_\mathrm{1}\cup G_\mathrm{2}}$ representing the sets of fast and slow generators. Wind, PV and storage units are represented by $w\in \mathcal{W}$, $m\in \mathcal{M}$ and $b\in \mathcal{B}$ respectively. {The uncertainties of renewable generation and demand in the microgrid are managed by the two-stage decision process. The unit commitment decisions are made in the first stage except for the fast-start generators, before the uncertainty is realized. Once most uncertain inputs (demand and renewable generation) are realized, the power outputs of committed units as well as the fast-start generators are decided to meet the load \cite{Conejo_2010,4773152}. The two-stage decision process in power system operations makes it natural to formulate the scheduling problem as a multi-stage stochastic program.} Based on the stochastic multi-temporal method proposed in \cite{7370811Teng}, the stochastic scheduling problem can be formulated as follows.
\subsection{Objective Function} \label{sec:4.1}
The objective of the scheduling problem is to minimize the microgrid average operation cost for all scenarios ($\forall s \in \mathcal{S}$) along with considered time horizon $t\in \{0,1,...,T\}$:
\begin{equation}
\begin{split}
    \min \sum_{s\in \mathcal{S}} & \sum_{t\in T} \pi_s(\sum_{g\in \mathcal{G}}c_g^{SU}z_{t,s,g}+\Delta t(\sum_{g\in \mathcal{G_\mathrm{1}}}c_g^{R1}y_{t,s,g}+ \\ & \sum_{g\in \mathcal{G_\mathrm{2}}}c_g^{R2}p_{t,s,g}
     +\sum_{l\in \mathcal{L}}c^{VOLL}(p_{t,s,l}^{c}+(q_{t,s,l}^{c})^2)))
\end{split}
\end{equation}
where $\pi_s$ is the probability associted with scenario $s$; $c_g^{SU}$, $c_g^{R1}/c_g^{R2}$ and $c^{VOLL}$ refer to start-up costs, running costs of fixed/flexible generators and the value of lost load (VOLL); $z_{t,s,g}$ and $y_{t,s,g}$ are binary variables of generator $g$ at time step $t$ in scenario $s$ with $1/0$ indicating starting up/not and on/off; $p_{t,s,g}$ and $p_{t,s,l}^{c}/q_{t,s,l}^{c}$ denote the active power produced by generators and active/reactive load shedding. 
\subsection{Constraints} \label{sec:4.2}
The traditional microgrid scheduling constraints that related to the generator operation, wind/PV curtailment and load shedding are omitted here. \cite{DLR} can be referred for more details. 
\subsubsection{Constraints of battery storage system}
\begin{subequations}
\begin{align}
    \label{p_b}
    & \eqref{Hv_lim},\,\eqref{Pc_lim}, \;\;\;\;\; \forall t,s,b\\
    \label{soc_cal}
    &\mathrm{SoC}_{t,s,b}E_{c,b} = \mathrm{SoC}_{t-1,s,b}E_{c,b} + \eta_b p_{t,s,b}\Delta t, \;\;\;\;\; \forall t,s,b\\
    \label{soc_lim}
    & \mathrm{SoC}_{\mathrm{min}} \le \mathrm{SoC}_{t,s,b} \le \mathrm{SoC}_{\mathrm{max}}, \;\;\;\;\; \forall t,s,b\\
    \label{SoC_T}
    & \mathrm{SoC}_{0,s,b} = \mathrm{SoC}_{T,s,b}, \;\;\;\;\; \forall t,s,b.
\end{align}
\end{subequations}
The power injection from the battery storage system to the microgrid is confined in \eqref{p_b} by the upper bound of the charging and discharging rate with $p_{b}$ in the original equations being replaced by $p_{t,s,b}$. The battery state of charge is quantified by \eqref{soc_cal} with the charging/discharging efficiency $\eta_b$. \eqref{soc_lim} imposes the upper and lower limits on the SoC of the storage devices. The SoC at the end of the considered time horizon is set to be a pre-specified value being equal to its initial value as in \eqref{SoC_T}.
    
\subsubsection{Constraints of AC power flow and power balance}
\begin{subequations}
\begin{align}
    & W_{t,s,ij}W_{t,s,ij}^* \le W_{t,s,ii}W_{t,s,jj}, \;\;\;\;\; \forall t,s,i,j \label{5-}\\
    & V_{\mathrm{min},i}^2\le W_{t,s,ii} \le V_{\mathrm{max},i}^2, \;\;\;\;\; \forall t,s,i \label{-5}\\
    & p_{t,s,i}^G = \sum_{\Omega_{g-i}} p_{t,s,g} + \sum_{\Omega_{w-i}} p_{t,s,w} \nonumber\\
    &  \quad\quad\quad\quad\quad + \sum_{\Omega_{m-i}} p_{t,s,m} + \sum_{\Omega_{b-i}} p_{t,s,b}, \;\;\;\;\; \forall t,s,i \label{6-}\\
    & q_{t,s,i}^G = \sum_{\Omega_{g-i}} q_{t,s,g} + \sum_{\Omega_{w-i}} q_{t,s,w} \nonumber\\
    & \quad\quad\quad\quad\quad + \sum_{\Omega_{m-i}} q_{t,s,m} + \sum_{\Omega_{b-i}} q_{t,s,b}, \;\;\;\;\; \forall t,s,i \label{-6}\\
    & p_{t,s,i}^D = \sum_{\Omega_{l-i}} p_{t,s,l} - \sum_{\Omega_{l-i}} p_{t,s,l}^c, \;\;\;\;\; \forall t,s,i \label{7-}\\
    & q_{t,s,i}^D = \sum_{\Omega_{l-i}} q_{t,s,l} - \sum_{\Omega_{l-i}} q_{t,s,l}^c, \;\;\;\;\; \forall t,s,i \label{-7}\\
    & p_{t,s,i} = p_{t,s,i}^G - p_{t,s,i}^D, \;\;\;\;\; \forall t,s,i \label{8-}\\
    & q_{t,s,i} = q_{t,s,i}^G - q_{t,s,i}^D, \;\;\;\;\; \forall t,s,i\\
    & p_{t,s,i} = \sum_{ij\in\mathcal{R}}p_{t,s,ij}, \;\;\;\;\; \forall t,s,i\\
    & q_{t,s,i} = \sum_{ij\in\mathcal{R}}q_{t,s,ij} -\mathrm{Im} {(W_{t,s,ii}\mathrm{j} Y_{i,sh} )}, \;\;\;\;\; \forall t,s,i \label{-8}\\
    & p_{t,s,ij} + \mathrm{j} q_{t,s,ij} = W_{t,s,ii} {Y_{i,sh}^{*}} \nonumber \\
    & \quad\quad\quad\quad\quad -  (W_{t,s,ii}-W_{t,s,ij})y_{ij}^{*}, \;\;\;\;\; \forall ij\in \mathcal{R}, t,s \label{9}\\
    & p_{t,s,ij}^2 + q_{t,s,ij}^2 \le S_{\mathrm{max},ij}^2, \;\;\;\;\; \forall ij\in \mathcal{R}, t,s \label{10}
\end{align}
\end{subequations}
Equations \eqref{5-} and \eqref{-5} are second-order cone constraints of voltages \cite{taylor_2015} where $W_{t,s,ij} = V_{t,s,i}V_{t,s,j}^*, \, \forall t,s,i,j$; $V_{t,s,i}/V_{t,s,j}$ are voltages at bus $i/j$ and $V_{\mathrm{min},i}/V_{\mathrm{max},i}$ are minimum/maximum voltage at bus $i$. Total active/reactive power generation and load at each bus are defined in \eqref{6-}/\eqref{-6} and \eqref{7-}/\eqref{-7} with $\Omega_{g/w/m/b-i}$ and $\Omega_{l-i}$ being the set of synchronous/wind/PV/storage units and loads connected to bus $i$. Note that the imported power from the main grid, $P_{t,s,im}/Q_{t,s,im}$ is include in $\Omega_{g-i}$ for simplicity. Power balance at each bus is given by \eqref{8-} to \eqref{-8} where $p_{t,s,ij}/q_{t,s,ij}$ are active/reactive power flow from bus $i$ to $j$ and $ij \in \mathcal{R}$ is the set of branches; $Y_{i,sh} = Y_{j,sh}$ denotes shunt susceptances at both ends of the line. \eqref{9} and \eqref{10} are the power flow and line rating constraints.

\subsubsection{Frequency security constraints subsequent to islanding events}
According to the derivation in Section \ref{sec:3}, \eqref{nadir_soc}\eqref{z_n12_linear}\eqref{z_N}\eqref{nadir_linear_1}, \eqref{rocof_cstr} and \eqref{fss_cstr} are incorporated into the  microgrid scheduling model as the frequency nadir, RoCoF and steady-state  constraints. {Therefore, the optimal microgrid inertia which includes both conventional and synthetic one, the PFR and the equivalent loss of generation will be determined in the microgrid scheduling model to ensure the minimum operational cost while maintaining the frequency constraints.} 

\color{black}
Different operations are coordinated before, during and after islanding events to ensure the frequency security. Before the islanding event, the microgrid operates in grid-connected mode. All the operating points of the dispatchable units and the imported power from the maingrid are set optimally according to the results from the scheduling model as indicated by the blue area of the above figure. These settings help to ensure the frequency security by allocating proper reserve, system inertia and power exchange with the maingrid.
    
At the time instant of an islanding event, the power exchange with the maingrid becomes zero almost instantaneously leading to a step power disturbance to the microgrid. This event can be detected with negligible delays by monitoring the main breaker at PCC and measuring the RoCoF of the frequency \cite{8561183,8389907}. 
    
After that, different frequency services instructed by the scheduling results, begin to react automatically including the frequency response from SGs, the SI provision from IBGs and the noncritical load shedding in order to facilitate the frequency regulation. After around tens of seconds, the frequency reaches and remains at the steady-state value until the recovery and restoration processes. Note that the fault repair and microgrid recovery and restoration processes are out of the scope of the proposed model.
\color{black}


\section{Case Studies} \label{sec:5}
\begin{figure}[!b]
    \centering
    \vspace{-0.35cm}
	\scalebox{0.35}{\includegraphics[]{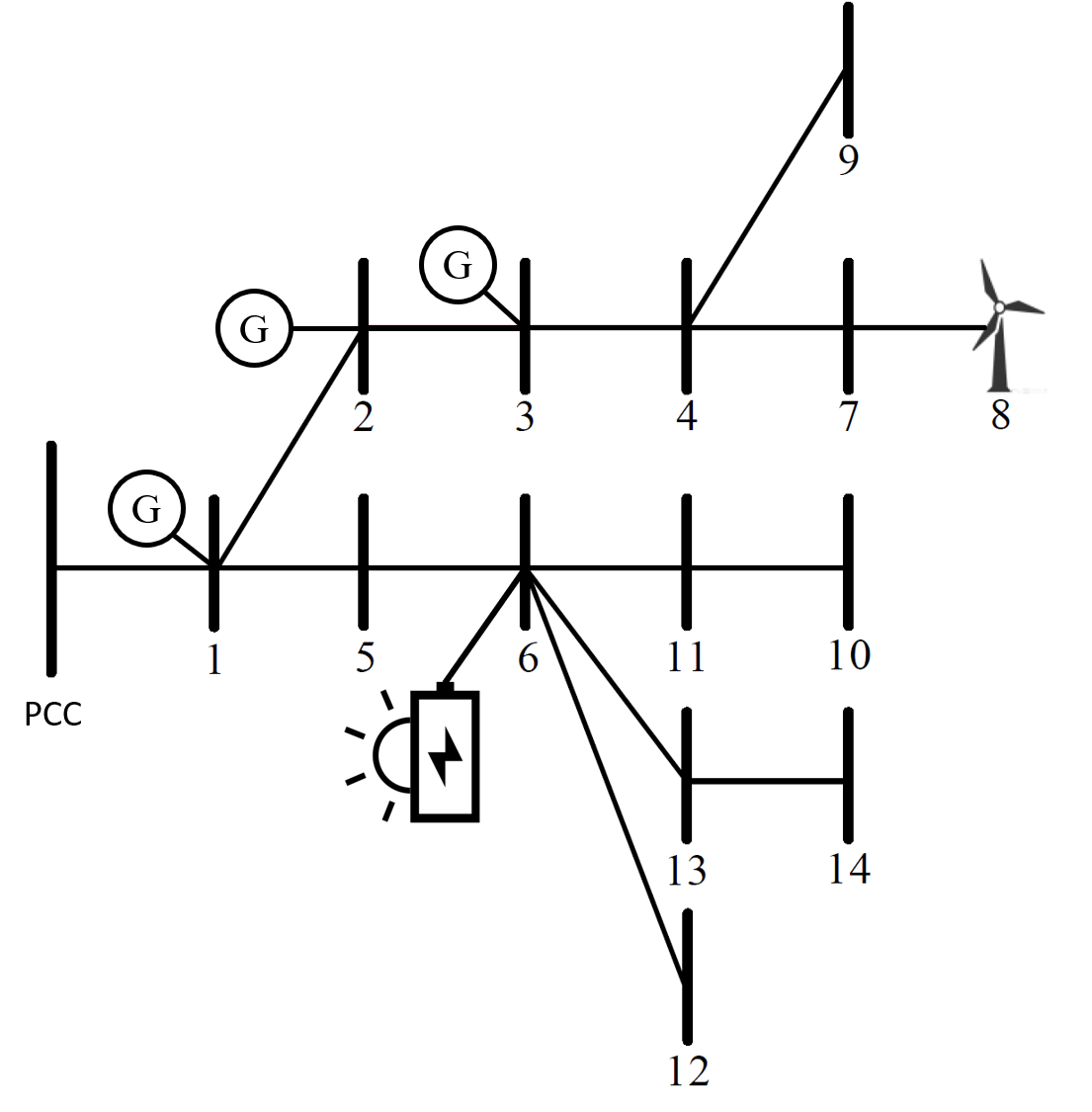}}
    \caption{\label{fig:14-bus}Modified 14-bus microgrid test system.}
    \vspace{-0.35cm}
\end{figure}

In order to demonstrate the performance of the proposed distributionally robust chance constrained microgrid scheduling model, case studies are carried out through the modified IEEE 14-bus distribution system \cite{14bus} as shown in Fig.~\ref{fig:14-bus}. {The optimization problem is solved in a horizon of 24 hours with the time step being 1 hour.} System parameters are set as follows: load demand $P_D\in [160,300]\,\mathrm{MW}$, damping $D = 0.5\% P_D / 1\,\mathrm{Hz}$, PFR delivery time $T_d = 10\,\mathrm{s}$. The frequency limits of nadir, steady-state value and RoCoF are set as: $\Delta f_\mathrm{lim} = 0.8\,\mathrm{Hz}$, $\Delta f_\mathrm{lim}^\mathrm{ss} = 0.5\,\mathrm{Hz}$ and $\Delta \dot f_\mathrm{lim} = 0.5\,\mathrm{Hz/s}$. Dispatchable SGs are installed at Bus 1,2 and 3 with a total capacity of $240\,\mathrm{MW}$. The PV-storage system and wind turbines locate at Bus 6 and 8 respectively. The parameters of battery devices are listed in Table~\ref{tab:battery}. The weather conditions are obtained from online numerical weather prediction \cite{weather}. The MISOCP-base optimization problem is solved by Gurobi (8.1.0) on a PC with Intel(R) Core(TM) i7-7820X CPU @ 3.60GHz and RAM of 64 GB.
\begin{table}[!t]
\renewcommand{\arraystretch}{1.2}
\caption{Parameters of Battery Storage Devices}
\label{tab:battery}
\noindent
\centering
    \begin{minipage}{\linewidth} 
    \renewcommand\footnoterule{\vspace*{-5pt}} 
    \begin{center}
        \begin{tabular}{ c | c | c | c | c }
            \toprule
             $\boldsymbol{\mathrm{SoC}_{\mathrm{min}}}$ &$\boldsymbol{\mathrm{SoC}_{\mathrm{max}}}$ &$\boldsymbol{\eta_b}$ &$\boldsymbol{|\bar P^{(d)ch}|\,\mathrm{[MW]}}$  & $\boldsymbol{E_c\,\mathrm{[MWh]}}$ \\ 
            \cline{1-5}
            $15\%$ & $85\%$  & $0.9$  & $50$ & $150$ \\ 
           \bottomrule
        \end{tabular}
        \end{center}
    \end{minipage}
\end{table} 
\subsection{Frequency nadir validation}
\begin{figure}[!t]
    \centering
    \vspace{-0.35cm}
	\scalebox{1.2}{\includegraphics[]{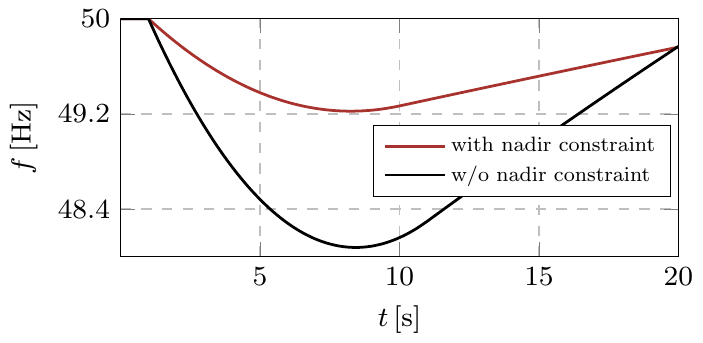}}
    \caption{\label{fig:f}{Microgrid frequency evaluation after an islanding event.}}
\end{figure}

\begin{figure}[!b]
    \centering
    \vspace{-0.35cm}
	\scalebox{0.46}{\includegraphics[]{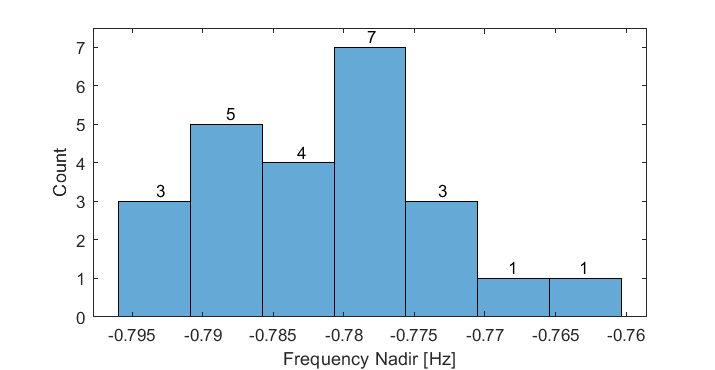}}
	\caption{\label{fig:Hitogram} Histogram of frequency nadirs in 24-hour scheduling.}
\end{figure}

\begin{figure}[!t]
    \centering
    \vspace{-0.35cm}
	\scalebox{1.2}{\includegraphics[]{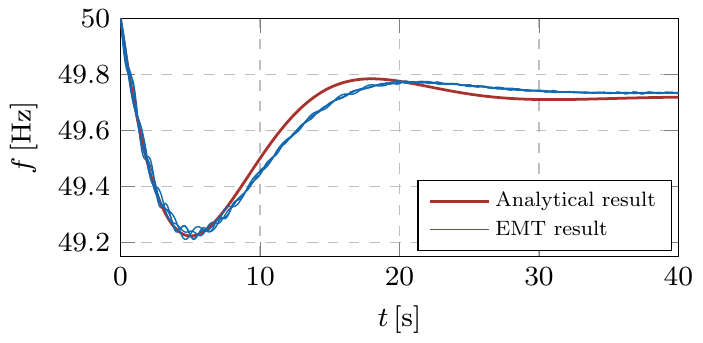}}
    \caption{\label{fig:f_EMT}{Frequency comparison: Analytical and EMT results.}}
\end{figure}
The approximation of the frequency nadir constraint discussed in Section \ref{sec:3.1} is assessed through dynamic simulation via Matlab/Simulink. {The microgrid frequency evaluation subsequent to an islanding event with and without (w/o) the nadir constraint is illustrated in Fig.~\ref{fig:f}.} The system operating conditions are selected at an arbitrary hour based on the optimal scheduled results: $P_D=162.7\,\mathrm{MW}$, $R = 50.1\,\mathrm{MW}$, $\Delta P_L= 37.0\,\mathrm{MW}$, $H = 86.0\,\mathrm{MWs/Hz}$. {It can be observed that the microgrid frequency decreases dramatically after an islanding event if the nadir constraint is not implemented. Even though the steady-state frequency is within the limit, the RoCoF and nadir constraints are violated. On the contrary, once the nadir constraint is considered in the microgrid scheduling model, all the frequency constraints can be maintained. The frequency nadir of $-0.77\,\mathrm{Hz}$ shows a good approximation yet conservativeness.} All the other conditions present a similar frequency evaluation trend thus not being covered. \textcolor{black}{Instead, to demonstrate the robustness of the proposed method in terms of effectiveness of the nadir constraints, the frequency nadir in each hour of the one-day scheduling if an unintentional islanding event occur is obtained through the dynamic simulation with the results depicted in Fig.~\ref{fig:Hitogram}. It is observed from the histogram that all the frequency nadirs during the 24-hour scheduling are close to the boundary (-0.8 $\mathrm{Hz}$) with the mean and standard deviation being $-0.7814\,\mathrm{Hz}$ and $0.008\,\mathrm{Hz}$ respectively, indicating a good robustness of the proposed method.}

\textcolor{black}{Additionally, the developed model is incorporated into the detailed EMT simulation and analyzed for the test case with $P_D=199.6\,\mathrm{MW}$, $R = 57.0\,\mathrm{MW}$, $\Delta P_L= 30.2\,\mathrm{MW}$, $H = 48.7\,\mathrm{MWs/Hz}$. As shown in Fig.~\ref{fig:f_EMT}, an unintentional islanding event occurs at $t=0\,\mathrm{s}$. The analytical result represent the Center-of-Inertia (CoI) frequency of the microgrid, whereas the 4 trajectories in the EMT result represent the local frequencies at the generation buses (Bus 1, 2, 6 and 8). Note that the SG at Bus 3 is not online in this hour. High oscillations depicted in the figure reflect the complexity of the EMT model at hand, as well as the level of controller interaction characteristic of low inertia system. It is observed that the frequency constraints in both cases can be maintained and the EMT result stays close to the analytical one despite a little mismatch after the frequency nadir, which is due to the approximation of the SG model in the analytical derivation. For more detailed SG, VSM and WT models, \cite{8579100,9066910} can be referred.}

\subsection{Impact of islanding frequency constraints and SI from RESs}
\begin{figure}[!t]
    \centering
    \vspace{-0.35cm}
	\scalebox{1.2}{\includegraphics[]{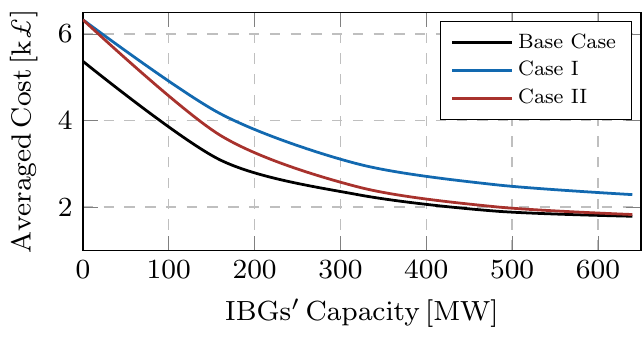}}
    \caption{\label{fig:IBG_C}Averaged cost at different operating conditions.}
\end{figure}

In this section, the influence of the frequency constraints subsequent to microgrid islanding events as well as the value of SI are investigated. System operation cost at different scheduling conditions is presented in Fig.~\ref{fig:IBG_C} with the cases defined as follows.
\begin{itemize}
  \item Base Case: Do not consider frequency dynamic constraints.
  \item Case I: Consider frequency dynamic constraints, and SI is not allowed.
  \item Case II: Consider frequency dynamic constraints, and SI is allowed.
\end{itemize}

\begin{figure}[!b]
    \centering
    \vspace{-0.35cm}
	\scalebox{1.2}{\includegraphics[]{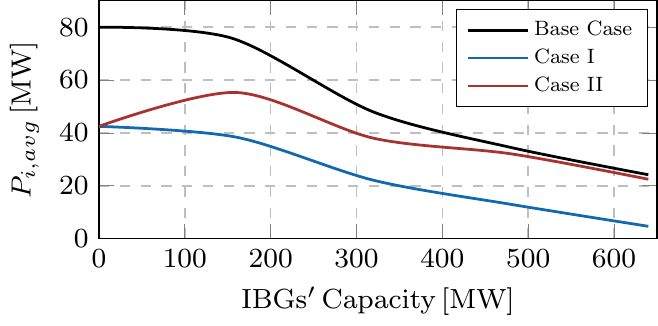}}
    \caption{\label{fig:P_in}Microgrid imported power at various IBGs' capacity.}
    \vspace{-0.35cm}
\end{figure}

Notably, the IBGs' capacity refers to the total capacity of wind turbines ($P_W^C$) and PV systems ($P_M^C$) with $P_W^C/P_M^C=3/5$; To avoid the PV and wind power curtailment due to the battery storage saturation, the total battery capacity (${\bar P^{(d)ch}},\,E_c$) also varies with the PV capacity, i.e., $\bar P^{(d)ch}:E_c:P_M^C = 1:3:2$. It is observed that the averaged system operation cost over 24 hours decreases along with the increase of IBGs' capacity in the system for all the three cases as more energy is supplied by the RESs. In the base case, the system operation cost always has the smallest value since the frequency dynamic constraints are not considered. As a consequence, violations of RoCoF and nadir constraints would be inevitable, should islanding events happen. For instance, $\Delta \dot f_{\mathrm{max}}\in [-3.81,-1.70]\,\mathrm{Hz/s}$ with an average of $-2.53\,\mathrm{Hz/s}$ and $\Delta f_{\mathrm{max}}\in [-13.08,-6.89]\,\mathrm{Hz}$ with an average of $-9.44\,\mathrm{Hz}$ are observed at IBGs' capacity of 320 $\mathrm{MW}$. Once the frequency dynamic constraints are included and SI is not allowed, the averaged cost (blue curve) grows to maintain the frequency limits by dispatch more partially-loaded SGs in the system for inertia provision only. 
The SI provision from RESs (Case II) reduces the operational cost significantly. This cost saving (the difference between Case I and II) becomes more obvious at high IGB's capacity where the cost of Case II is almost the same as that of the base case, which highlights the effectiveness and value of SI provision especially in high PE-penetrated microgrids. {The computational time of each optimization in different cases varies between $[59.14, 169.84]\,\mathrm{s}$ with an average of $100.92\,\mathrm{s}$.}

The averaged imported power from the main grid, $P_{i,avg}$ is also depicted in Fig.~\ref{fig:P_in}. In the Base Case, the imported power starts to decrease after IBGs' capacity becomes higher than about $110\,\mathrm{MW}$ since less energy is needed from the main grid. If the frequency dynamic constraints are considered (Case I), the microgrid cannot deal with the large disturbance without the SI. Therefore, $P_{i,avg}$ is reduced by around a half compared to Base Case in order to decrease the system disturbance level. In Case II, the system available SI becomes higher as the IBGs' capacity rises. Therefore, the microgrid can withstand larger disturbance without violating the frequency constraints, thus enabling more imported power from the main grid compared to Case I, which also justifies the cost saving in Fig.~\ref{fig:IBG_C}. 

\begin{figure}[!t]
    \centering
    \vspace{-0.35cm}
	\scalebox{1.1}{\includegraphics[]{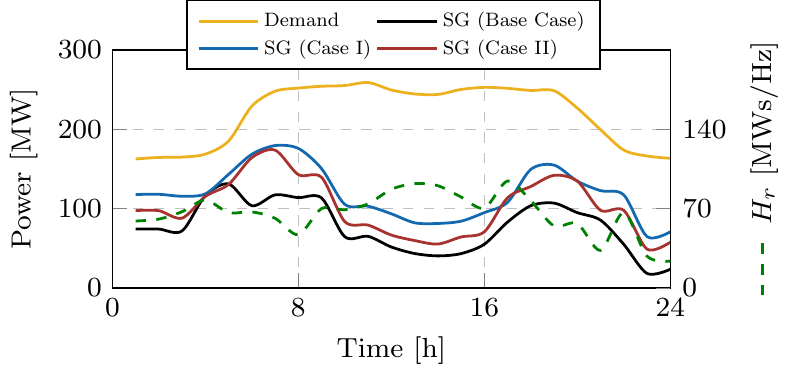}}
    \caption{\label{fig:disp}Microgrid scheduling results within one day.}
\end{figure}

The effects on the system SGs' dispatch in different cases are investigated as well with the results of 24 hours shown in Fig.~\ref{fig:disp} together with the demand and total SI ($H_r$) profile. The IBGs' capacity in the system is $160\,\mathrm{MW}$. The implementation of the frequency dynamic constraints (Case I) induces more power dispatched from SGs in almost all hours compared to Base Case such that the power from the main grid could be reduced. With the SI from RESs (Case II), more power can be supplied by the main grid and RESs leading to a declined SG power. In addition, the total SI from RESs, varying in the range of $[23-94]\,\mathrm{MWs/Hz}$ is also plotted, where its inverse relationship with SG power in Case II is observed. In particular, during the time of low net demand (i.e., $t\in[12,16]\,\mathrm{h}$), a significant amount of SI is scheduled from RESs due to the lower inertia from SGs and vice versa.

\subsection{Impact of uncertainty level of demand shedding during islanding events}
\begin{figure}[!b]
    \centering
    \vspace{-0.35cm}
	\scalebox{1.2}{\includegraphics[]{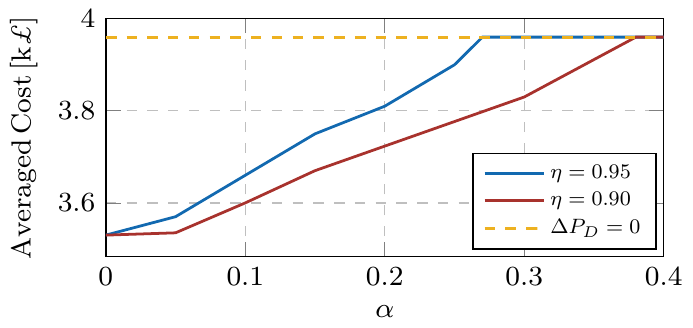}}
    \caption{\label{fig:sigma}Averaged operation cost at different uncertainty levels.}
    \vspace{-0.35cm}
\end{figure}
In order to maintain the frequency constraints during unintentional islanding events, noncritical load shedding is implemented to reduce the disturbance magnitude. {The IBGs' total capacity is set to be $160\mathrm{MW}$.} The uncertainty level $\alpha$ associated with the noncritical load shedding as defined in \eqref{alpha} is evaluated in this section. Its influence on the averaged microgrid operation cost during 24 hours is depicted in Fig.~\ref{fig:sigma} with different confidence levels ($\eta = 0.95$ and $0.90$). As expected, a higher uncertainty level generates more operational cost since its effect on reducing the disturbance becomes less reliable. Moreover, one can find that as the confidence level is reduced from $0.95$ to $0.90$, the cost decreases by approximate $10\%$, which highlights that the trade-off between the risk level and microgrid operation cost needs to be well balanced. It is also worth noticing that as the uncertainty level increases above some value (0.27 or 0.38), the microgrid operation cost becomes the same as the case where noncritical load shedding is not implemented, represented by the dashed yellow curve. Since the Chebyshev inequality is used in the deviation of nadir constraints \eqref{nadir_nonlinear}, which gives the lower bound regardless of the actual distribution of the noncritical load shedding, conservative results are obtained. Therefore, if this is the case in practice, system operators should either pursue more knowledge of the load shedding distribution or decrease the confidence level to achieve benefits in terms of microgrid operation cost saving.

\section{Conclusion and Future Work}    \label{sec:6}
This paper proposes a novel microgrid scheduling model enabling optimal selection of the SI from IBGs while maintaining a minimum operational cost and frequency constraints subsequent to an islanding event. Based on detailed microgrid frequency dynamics and the state-of-art SI control schemes of the IBGs, the frequency metrics subjected to an islanding event, which is modeled as a step disturbance, are derived analytically. The uncertainty level associated with the noncritical load shedding is modeled via an ambiguity set without the knowledge of its specific distribution, leading to a distributionally robust reformulation of the frequency constraints given a certain confidence level. The nonlinear and nonconvex nadir constraint is approximated by SOC relationship with the conservativeness and accuracy being illustrated. An overall MISOCP-based optimization problem is formulated and can be solved effectively using commercial solvers. Case studies demonstrate the importance and necessity of islanding event consideration and the value of SI provision from IBGs in terms of microgrid operation cost saving. The impact of the uncertainty level of the noncritical load shedding is also investigated.

The proposed model can be enhanced in several directions. Instead of assuming the noncritical load can be shedding simultaneously, at the time instant when the islanding event occurs, time delays in different types of load due to the event detection and communications can be modeled in more detail. In addition, the emulated damping from RESs can also be considered, providing more degrees of freedom in the frequency regulation and microgrid scheduling.
\bibliographystyle{IEEEtran}
\bibliography{bibliography}




\vfill

\enlargethispage{-50mm}

\end{document}


\begin{varwidth}{\linewidth}

\begin{tikzpicture}
\begin{axis}[
    scaled ticks=false,
    tick label style={/pgf/number format/fixed},
    colormap name=viridis,
    width=7.25cm,
    height=4cm,
    xlabel={Installed Wind Capacity  [GW]},
    ylabel={Operational Cost [B\pounds]},
    xmin=0, xmax=60,
    ymin=7, ymax=16,
    xtick={0,10,20,30,40,50,60},
    xmajorgrids=true,
    ymajorgrids=true,
    legend style={at={(axis cs:59.5,15.8)},anchor=north east,nodes={scale=0.75, transform shape}},
    legend cell align={left},
    grid style=dashed,
]
\footnotesize
\addplot[
    smooth,
    thick,
    color=pBlack,
    ]
    table {data/SI_Value/data0.txt};    
    \addlegendentry{\footnotesize w/o FC, w/o SI}       
    
\addplot[
    smooth,
    thick,
    color=pBlue,
    ]
    table {data/SI_Value/data1.txt};        
    \addlegendentry{\footnotesize with FC, w/o SI}  
    
\addplot[
    smooth,
    thick,
    color=pRed,
    ]
    table {data/SI_Value/data2.txt};        
    \addlegendentry{\footnotesize with FC, with SI}  

\end{axis}

\end{tikzpicture} 

\end{varwidth}